\documentstyle[aps, pra, twocolumn, amsfonts]{revtex}

\def\qed{$\Box$}

\newtheorem{thm}{Theorem}
\newtheorem{cor}[thm]{Corollary}

\begin{document}

\title{On the Theory of Quantum Secret Sharing}

\author{Daniel Gottesman\thanks{e-mail: gottesma@microsoft.com}}

\address{Microsoft Corporation\\One Microsoft Way\\Redmond, WA
98052}

\maketitle

\begin{abstract}
I present a variety of results on the theory of quantum secret
sharing.  I show that any mixed state quantum secret sharing
scheme can be derived by discarding a share from a pure state
scheme, and that the size of each share in a quantum secret
sharing scheme must be at least as large as the size of the
secret.  I show that the only constraints on the existence of
quantum secret sharing schemes with general access structures are
monotonicity (if a set is authorized, so are larger sets) and the
no-cloning theorem.  I also discuss some aspects of sharing
classical secrets using quantum states.  In this situation, the
size of each share can sometimes be half the size of the classical
secret.
\end{abstract}

\pacs{03.67.Dd, 03.67.-a}

\section{Introduction}

In a classical secret sharing scheme, some sensitive classical
data is distributed among a number of people such that certain
sufficiently large sets of people can access the data, but smaller
sets can gain no information about the shared secret.  For
instance, a possible application is to share the key for a joint
checking account shared by many people.  No individual is able to
withdraw money, but sufficiently large groups can use the account.

One particularly symmetric variety of secret sharing scheme is
called a {\em threshold scheme}.  A $(k, n)$ classical threshold
scheme has $n$ shares, of which {\em any} $k$ are sufficient to
reconstruct the secret, while any set of $k-1$ or fewer shares has
no information about the secret.  Blakely~\cite{Blakely} and
Shamir~\cite{Shamir} showed that threshold schemes exist for all
values of $k$ and $n$ with $n \ge k$.

It is also possible to consider more general secret sharing
schemes which have an asymmetry between the power of the different
shares.  For instance, one might consider a scheme with four
shares $A$, $B$, $C$, and $D$.  Any set containing $A$, $B$, and
$C$ {\em or} $A$ and $D$ can reconstruct the secret, but any other
set of shares has no information.  In this example, the presence
of $A$ is essential to reconstructing the secret, but not
sufficient --- $A$ needs the help of either $D$ or both $B$ and
$C$.  This particular scheme can be constructed by taking a $(5,
7)$ threshold scheme, and assigning 3 shares to $A$, 2 to $D$, and
1 to each of $B$ and $C$, but other schemes exist which cannot be
constructed by bundling together shares of a threshold scheme. The
list of which sets are able to reconstruct the secret is called an
{\em access structure} for the secret sharing scheme.  It turns
out that a secret sharing scheme exists for any access structure,
provided it is monotone~\cite{AllCSS} --- i.e., that if a set $S$
can reconstruct the secret, so can all sets containing $S$.

With the advent of quantum computation, it is possible that
quantum information may someday be as commonplace as classical
information, and we may wish to protect it the same ways as we
protect classical information.  Using quantum secret
sharing~\cite{QSS}, we could perhaps create joint checking
accounts containing quantum money~\cite{WiesnerQM}, or share
hard-to-create ancilla states~\cite{GC}, or perform a secure
distributed quantum computation.  \cite{QSS}~showed some basic
results about quantum secret sharing schemes, including the
existence of quantum threshold schemes.  A quantum $((k, n))$
threshold scheme (the use of double parentheses distinguishes it
from a classical scheme) exists provided the no-cloning theorem is
satisfied --- i.e., $n/2 < k \le n$.  In this paper, I will prove
some further results about quantum secret sharing schemes with
general access structures, including the fact that the no-cloning
theorem and monotonicity provide the only restriction on the
existence of quantum secret sharing schemes.

Another possible application of quantum states to secret sharing
is to create secret sharing schemes sharing classical data using
quantum states~\cite{HBB,KKI}.  This could allow, for instance,
for more secure distribution of the shares of the scheme.  I will
show below that it can also produce more efficient schemes: in any
purely classical scheme, the size of each important share must be
at least as large as the size of the secret, whereas using quantum
states to share a classical secret, we can sometimes make each
share {\em half} the size of the secret.

In the theory of classical secret sharing, one sometimes considers
schemes which do not completely hide the secret from unauthorized
groups of people, or from which the secret cannot be perfectly
reconstructed even by authorized sets.  I will not consider the
quantum generalizations of such schemes.  I will only consider the
theory of perfect secret sharing schemes, in which the data is
either completely revealed or completely hidden, with no middle
ground.

\section{Quantum Secret Sharing}

I will begin by reviewing some results from~\cite{QSS} which will
form the basis of much of the later discussion.  In a perfect
quantum secret sharing scheme, any set of shares is either an {\em
authorized} set, in which case someone holding all of those shares
can exactly reconstruct the original secret, or an {\em
unauthorized} set, in which case someone holding just those shares
can acquire no information at all about the secret quantum state
(that is, the density matrix of an unauthorized set is the same
for all encoded states). For a generic state split up into a
number of shares, most sets will be neither authorized nor
unauthorized --- quantum secret sharing schemes form a special set
of states.

One constraint on quantum secret sharing schemes is an obvious one
inherited from classical schemes.  Any secret sharing scheme must
be {\em monotonic}.  That is, if we increase the size of a set, it
cannot switch from authorized to unauthorized (the indicator
function which is 0 for unauthorized sets and 1 for authorized
sets is monotonic).

As we shall see in section~\ref{sec:general}, the only other
constraint on quantum secret sharing schemes is the no-cloning
theorem~\cite{WZ,Dieks}.  We cannot make two copies of an unknown
quantum state.  Therefore, we cannot distribute the shares of
quantum secret sharing scheme into two disjoint authorized sets
(each of which could produce a copy of the original state).  Since
every set is either authorized or unauthorized, this implies the
complement of an authorized set is always an unauthorized set.

A {\em pure state} quantum secret sharing scheme encodes pure
state secrets as pure states (when all of the shares are
available).  A {\em mixed state} quantum secret sharing scheme may
encode some or all pure states of the secret as mixed states.
Pure state schemes have some special properties, as a consequence
of the following theorem, but the general quantum secret sharing
scheme is a mixed state scheme.

Theorem~\ref{thm:QSScond} and Corollary~\ref{thm:pure} first
appeared in~\cite{QSS}.

\begin{thm}
\label{thm:QSScond} Let ${\cal C}$ be a subspace of a Hilbert
space ${\cal H}$ which can be written as tensor product of the
Hilbert spaces of various coordinates. Then ${\cal C}$ corrects
erasure errors\footnote{An erasure error is a general error on a
known coordinate.  For instance, it replaces the coordinate with a
state $|e\rangle$ orthogonal to the regular Hilbert space.  Recall
that a quantum error-correcting code of distance $d$ can correct
$d-1$ erasure errors or $\lfloor (d-1)/2 \rfloor$ general errors.}
on a set $K$ of coordinates iff
\begin{equation}
\langle \phi | E | \phi \rangle = c(E) \label{eq:erasure}
\end{equation}
(independent of $|\phi\rangle \in {\cal C}$) for all operators $E$
acting on $K$.  A pure state encoding of a quantum secret is a
quantum secret sharing scheme iff the encoded space corrects
erasure errors on unauthorized sets and it corrects erasure errors
on the complements of authorized sets.
\end{thm}

\bigskip \noindent
{\bf Proof:} The first equivalence follows from the theory of
quantum error-correcting codes.  To recover the original secret on
an authorized set, we must be able to compensate for the absence
of the remaining shares, which is to say from an erasure error on
the complement of the authorized set.  The
condition~(\ref{eq:erasure}) implies that measuring any Hermitian
operator on the coordinates $K$ gives us no information about
which state in ${\cal C}$ we have.  This means the density matrix
on $K$ does not depend on the state, which is precisely the
condition we need an unauthorized set to satisfy. \hfill \qed
\bigskip

As a corollary, we find that pure state schemes are only possible
for a highly restricted class of access structures.

\begin{cor}
\label{thm:pure} In a pure state quantum secret sharing scheme,
the authorized sets are precisely the complements of the
unauthorized sets.
\end{cor}

\noindent
{\bf Proof:}  By the no-cloning theorem, the complement
of an authorized set is always an unauthorized set.  By
theorem~\ref{thm:QSScond}, for a pure state scheme, we can correct
erasure errors on any unauthorized set.  This means we can
reconstruct the secret in the absence of those shares; that is,
the complement is an authorized set. \hfill \qed
\bigskip

Suppose we start with an arbitrary quantum access structure (a set
of authorized sets) and add new authorized sets, filling out the
result to be monotonic.  For instance, if we started with the
access structure $ABC$ or $AD$ from the introduction (any set
containing $A$, $B$, and $C$ is authorized, as is any set
containing both $A$ and $D$), we could add the set $BD$ (so any
set containing $B$ and $D$ is also now authorized).  We wish to
continue to satisfy the no-cloning theorem as well, so we never
add a new authorized set contained in the complement of an
existing authorized set.  This ensures that the complement of
every authorized set remains an unauthorized set.  For instance,
in the example, we could not have added $BC$ as an authorized set,
since its complement $AD$ is already authorized.

Initially, there may be unauthorized sets whose complements are
also unauthorized, but if we continue adding authorized sets, we
will eventually reach a point where the authorized and
unauthorized sets are always complements of each other, as is
required for a pure state scheme.  In the example, we could add
$CD$ as an authorized set.  Now, the authorized sets are all sets
containing $ABC$, $AD$, $BD$, or $CD$.  At this point, we will
have to stop adding authorized sets --- any more would violate the
no-cloning theorem.  Thus, an access structure where the
authorized and unauthorized sets are complements of each other is
a {\em maximal} quantum access structure.

Pure state schemes and maximal access structures may seem like a
very special situation, but in fact they play a central role in
the theory of quantum secret sharing because of the following
theorem:

\begin{thm}
\label{thm:mixed} Every mixed state quantum secret sharing scheme
can be described as a pure state quantum secret sharing scheme
with one share discarded.  The access structure of the pure state
scheme is unique.
\end{thm}

\noindent
{\bf Proof:} Given a superoperator that maps the Hilbert
space ${\cal S}$ of the secret to density operators on ${\cal H}$
(which is a tensor product of the Hilbert spaces of the various
shares), we can extend the superoperator to a unitary map from
${\cal S}$ to ${\cal H} \otimes {\cal E}$ for some space ${\cal
E}$.  We assign this additional Hilbert space to the extra share.
In other words, we can ``purify'' the mixed state encoding by
adding an extra share.  The original mixed state scheme is
produced by discarding the extra share.  I claim that the new pure
state encoding is a quantum secret sharing scheme.

Sets on the original shares remain authorized or unauthorized, as
they were before adding ${\cal E}$.  Given a set $T$ including the
extra share, look at the complement of $T$, which is a set not
including ${\cal E}$ and is thus either authorized or unauthorized
(in the new scheme as well as the old).  For instance, if we
purify the scheme ($ABC$ or $AD$) by adding a fifth share $E$, the
complement of $CDE$ is unauthorized, while the complement of $DE$
is authorized.  If the complement is authorized, then we can
correct for erasures on $T$, and condition~(\ref{eq:erasure})
holds for $T$ --- we can get no information about the secret from
$T$, and $T$ is unauthorized. If the complement of $T$ is
unauthorized, we can correct erasures on the complement.
Therefore, we can reconstruct the state with just $T$, and $T$ is
authorized. Thus, the new scheme is secret sharing.

It is clear from the argument that any other purification of the
mixed state scheme would produce the same access structure. \hfill
\qed
\bigskip

In~\cite{QSS}, we presented a class of quantum secret sharing
schemes where every share had the same size as the secret.  One
might wonder if it is possible to do better.  For instance, can we
make one share much smaller than the secret, possibly at the cost
of enlarging another share?  The answer is no, provided we only
consider important shares (unimportant shares never make a
difference as to whether a set is authorized or unauthorized).

\begin{thm}
\label{thm:Qsize} The dimension of each important share of a
quantum secret sharing scheme must be at least as large as the
dimension of the secret.
\end{thm}

\noindent
{\bf Proof:} We need only prove the result for pure
state schemes. By theorem~\ref{thm:mixed}, the result for mixed
state schemes will follow.

Let $S$ be an important share in a pure state quantum secret
sharing scheme.  Then there is an unauthorized set $T$ such that
$T \cup \{S\}$ is authorized.  Share the state $|0\rangle$ and
give the shares of $T$ to Bob and the remaining shares (including
$S$) to Alice.  By corollary~\ref{thm:pure}, Alice's shares form
an authorized set; she can correct for erasures on $T$.  By
theorem~\ref{thm:write} below, this means Alice can perform any
operation she likes on the secret without disturbing Bob's shares.
She can equally well perform quantum interactions between the
secret and other quantum states held by her.  In particular, if
Alice has state $|\psi\rangle$ from a Hilbert space of dimension
$s$ (the size of the secret), she can coherently swap it into her
shares of the secret sharing scheme, which now encodes the state
$|\psi\rangle$. Then Alice sends just the share $S$ to Bob. Bob
now has an authorized set, so he can reconstruct $|\psi\rangle$.
Therefore, by theorem~\ref{thm:obvious} below, share $S$ must have
had dimension at least $s$ as well. \hfill \qed
\bigskip

The above proof depends on two theorems of interest outside the
theory of quantum secret sharing.  The first is obvious, and it is
also true; it has not, to my knowledge, appeared before in the
literature.

\begin{thm}
\label{thm:obvious}  Even in the presence of preexisting
entanglement, sending an arbitrary state from a Hilbert space of
dimension $s$ requires a channel of dimension $s$.
\end{thm}

\noindent {\bf Proof:} This proof is due to Michael
Nielsen~\cite{MNcom}.

Assume that in addition to whatever entanglement is given, Alice
and Bob share a cat state $\sum |i\rangle_A |i\rangle_B$ of
dimension $s$. Using a straightforward variant of superdense
coding~\cite{Wiesner}, Alice can encode one of $s^2$ classical
states in this cat state.  Now Alice transmits her half of the cat
state to Bob, using the preexisting entanglement if it helps.  Bob
can now reconstruct the classical state, so by the bounds on
superdense coding~\cite{CDNT}, Alice must have used a channel of
dimension $s$. \hfill \qed
\bigskip

The second theorem is more interesting.  It says that if Alice can
read a piece of quantum data, she can also change it any way she
likes, without disturbing any entanglement of the encoding with
the outside. There will be no way to tell that the data has been
changed.

\begin{thm}
\label{thm:write} Suppose a superoperator ${\cal S}$ maps a
Hilbert space $H$ to density operators on $A \otimes B$, and
${\cal S}$ restricted to $A$ (that is, traced over $B$) is
invertible (by quantum operation). Then for any unitary $U: H
\rightarrow H$, there exists a unitary operation $V: A \rightarrow
A$ such that $V \circ {\cal S} = {\cal S} \circ U$.
\end{thm}

\noindent {\bf Proof:} We can extend the superoperator ${\cal S}$
to a unitary operator $W$ and enlarge $B$ with the necessary extra
dimensions. If $V$ works for $W$, it will also work for ${\cal
S}$. Since $W$ is invertible on $A$, the image subspace corrects
erasure errors on $B$, and
\begin{equation}
\langle \psi | E | \psi \rangle = c(E)
\label{eq:structure}
\end{equation}
for any operator $E$ acting on $B$, where $c(E)$ is independent of
$|\psi\rangle \in W(H)$.  Choose a basis $|j\rangle_B$ for $B$.
Given any state $|\psi\rangle$ in the image of $W$, we can write
it as
\begin{equation}
|\psi\rangle = \sum |\psi_j\rangle_A |j\rangle_B.
\end{equation}
(The states $|\psi_j\rangle$ are not necessarily orthogonal,
although we could have made them orthogonal for any single
$|\psi\rangle$.)  If we let $E$ be a projection on the basis
states of $B$, or a projection on the basis states followed by a
permutation of those basis states, (\ref{eq:structure}) implies
that the inner products $\langle \psi_i | \psi_j \rangle$ are
independent of $|\psi\rangle$. Therefore, there is a unitary
operation $V$ acting on $A$ that takes any set of states
$|\psi_j\rangle_A$ for $|\psi\rangle \in W(H)$ to the set of
states $|\phi_j\rangle_A$ for any state $|\phi\rangle \in W(H)$.
In fact, $V$ will map $|\psi\rangle$ to $|\phi\rangle$.

More generally, and by the same logic, given any two bases of
$W(H)$, there will be a unitary $V$ on $A$ that takes one to the
other.  Given $U: H \rightarrow H$, we can define $U$ as mapping a
basis $|v_i\rangle$ to basis $|w_i\rangle$.  Then define $V: A
\rightarrow A$ as an operator that maps $W|v_i\rangle$ to
$W|w_i\rangle$, and the theorem follows.
\hfill \qed
\bigskip

I conclude this section with an easy theorem that will be needed
in the construction of a general access structure.

\begin{thm}
\label{thm:concatenation} If $S_1$ and $S_2$ are quantum secret
sharing schemes, then the scheme formed by concatenating them
(expanding each share of $S_1$ as the secret of $S_2$) is also
secret sharing.
\end{thm}

The reason this requires proof is that, due to some nonlocal
quantum effect, it might have been possible to get more
information from sets in two copies of $S_2$ than can be accessed
from just one of the sets.

\bigskip \noindent
{\bf Proof:}  By theorem~\ref{thm:mixed}, we need only consider
pure state schemes.  Then the concatenated scheme $S$ is a pure
state scheme too.  Suppose we have some set of shares $T$.  We can
write it as the union $\bigcup T_i$, where $T_i$ is a set on the
$i$th copy of $S_2$.  Consider the set $U$ of copies on which
$T_i$ is authorized.  $U$ is either an authorized or an
unauthorized set of $S_1$.  If it is authorized, then our big set
$T$ is certainly authorized --- we reconstruct the copies of $S_2$
in $U$, and use $U$ to reconstruct the original secret.

If $U$ is unauthorized, we look at the complement of $T$.  It can
be written as a union $\bigcup T'_i$, where $T'_i$ is the
complement of $T_i$ in its copy of $S_2$.  $T'_i$ is authorized
whenever $T_i$ is unauthorized.  Therefore, the set of copies on
which $T'_i$ is authorized is the complement of $U$, which is
authorized.  Thus, the complement of $T$ is authorized, so $T$ is
unauthorized.
\hfill \qed
\bigskip

Clearly the proof works equally well for more complicated
concatenation schemes, with multiple levels or with a different
scheme $S_2$ for each share of $S_1$.  Also note that if we bundle
shares together (assigning two or more shares to the same person),
the result is still a secret sharing scheme.

\section{Construction of a General Access Structure}
\label{sec:general}

This section will be devoted to proving that monotonicity and the
no-cloning theorem provide the only restrictions on the existence
of quantum secret sharing schemes.  The same result has been shown
by Adam Smith~\cite{AScom} by adapting a classical construction.
The construction given here is undoubtedly far from optimal in
terms of the share sizes of the resulting schemes.

\begin{thm}
\label{thm:general} A quantum secret sharing scheme exists for an
access structure $S$ iff $S$ is monotone and satisfies the
no-cloning theorem (i.e., the complement of an authorized set is
an unauthorized set).  For any maximal quantum access structure
$S$, a pure state scheme exists.
\end{thm}

It will be helpful to first understand an analogous classical
construction~\cite{AllCSS}.  Any access structure can be written
in a disjunctive normal form, which is the OR of a list of
authorized sets.  For our standard example, with authorized sets
$ABC$ and $AD$, the normal form is ($A$ AND $B$ AND $C$) OR ($A$
AND $D$). This normal form provides a construction in terms of
threshold schemes --- the AND gate corresponds to a $(2, 2)$
threshold scheme (which has one authorized set $A$ AND $B$), while
the OR gate corresponds to a $(1, 2)$ threshold scheme (for which
$A$ OR $B$ is authorized).  Then by concatenating the appropriate
set of threshold schemes, we get a construction for the original
access structure.

In the quantum case, this construction fails, because by the
no-cloning theorem, there is no $((1, 2))$ quantum threshold
scheme.  A single authorized set (such as $A$ AND $B$ AND $C$)
still corresponds to a quantum threshold scheme (a $((3,3))$
scheme in this case), but to take the OR of these authorized sets,
we will have to do something different.  We will replace the $((1,
2))$ scheme with $((r, 2r-1))$ schemes (which correspond to
majority functions instead of OR).  $r$ of the shares will be the
individual authorized sets of the desired access structure, and
the other $r-1$ shares will be from another access structure that
is easier to construct.

The full construction is recursive.  Given constructions of access
structures for $n-1$ shares, we will construct all {\em maximal}
access structures for $n$ shares.  From maximal access structures
on $n$ shares we will be able to construct all access structures
on $n$ shares.  We can start from the base case of 1 share, which
just has the trivial $((1, 1))$ access structure.  The
construction will assume we know how to create threshold schemes,
for instance using the construction in~\cite{QSS}.

Given any maximal access structure $S$ on $n$ shares, consider the
access structure $S'$ obtained by discarding one share.  Certainly
$S'$ is still monotonic and still satisfies the no-cloning
theorem.  Therefore, by the inductive hypothesis, we have a
construction for the access structure $S'$.  Now, following the
proof of theorem~\ref{thm:mixed}, add an additional share to $S'$
putting it in an overall pure state.  By the proof of
theorem~\ref{thm:mixed}, we know the resulting scheme is in fact a
quantum secret sharing scheme.  It is not hard to see that $S$ is
the unique access structure produced this way.

For instance, the maximal access structure $ABC$ OR $AD$ OR $BD$
OR $CD$ can be formed by purifying the (mixed state) scheme with
access structure $ABC$ (just a $((3, 3))$ threshold scheme).

Now suppose we are given a general quantum access structure $S$ on
$n$ shares.  We describe this access structure by a list of its
minimal authorized sets $A_1, A_2, \ldots, A_r$.  As mentioned
above, $A_i$ by itself defines a quantum access structure --- a
$((k, k))$ threshold scheme, in fact, if $A_i$ contains $k$
shares.

$S$ has a total of $r$ minimal authorized sets.  Let us take a
$((r, 2r-1))$ quantum threshold scheme, and expand each of its
shares using another secret sharing scheme.  Share $i$, for $i =
1, \ldots, r$, is expanded using the threshold scheme associated
with the set $A_i$. Shares $r+1$ through $2r-1$ will all be
expanded using another secret sharing scheme $S'$.

$S'$ will be a pure state scheme, with a maximal access structure
which can be achieved by adding authorized sets to $S$.  That
means when $A$ is an authorized set of $S$ (so it contains some
$A_i$), it is also an authorized set of $S'$.  Therefore, we can
reconstruct the last $r-1$ shares of the $((r, 2r-1))$ scheme, as
well as at least one of the first $r$ shares, so $A$ is an
authorized set for the concatenated scheme.

Conversely, if we have a set $B$ which does not include any of the
sets $A_i$, we do not have an authorized set for any of the
schemes $A_i$.  $B$ might be an authorized set for the scheme
$S'$, but that only gives us authorized sets for at most $r-1$
shares of the $((r, 2r-1))$ scheme.  Therefore, $B$ is an
unauthorized set.  This shows that the access structure of the
concatenated scheme is exactly $S$, completing the construction.

As an example, consider this construction applied to the access
structure $ABC$ OR $AD$.  The three rows represent shares of a
$((2, 3))$ scheme, so authorized sets on any two rows suffice to
reconstruct the secret.  Repeated letters imply bundling, so $A$
gets a share from each of the first two rows, as well as one from
the third row.
\begin{equation}
((2, 3))\ {\rm scheme} \left\{
\begin{array}{l}
((3, 3)):\ A,\ B,\ C \\
((2, 2)):\ A,\ D \\
S'
\end{array}
\right.
\end{equation}
The first two rows are threshold schemes.  $S'$ is a maximal
access structure containing $\{A, B, C\}$ and $\{A, D\}$.  For
instance, in this case, $S'$ could be the scheme $ABC$ OR $AD$ OR
$BD$ OR $CD$ which we constructed earlier; or we could just use
the trivial scheme with authorized set $\{A\}$ (give $A$ the
secret).

I noted in the introduction that this particular scheme can be
easily constructed directly from a $((5, 7))$ threshold scheme.
However, not all access structures can be made by bundling
together shares of a threshold scheme (for instance, $ABCD$ OR
$ADE$ OR $BCD$ cannot be so constructed\footnote{For quantum
access structures, threshold schemes suffice for fewer than five
shares, whereas for classical access structures, there are
examples where they fail for four shares. This is because the
four-share classical examples would violate the no-cloning
theorem.} --- $E$ would have to get more shares of the threshold
scheme than $B$ since $ADE$ is authorized while $ABD$ is not, but
$BCD$ is authorized while $CDE$ is not), while the recursive
construction always works.

\section{Sharing Classical Secrets}

We can also use quantum states to share classical secrets, a
process previously considered in \cite{HBB} and \cite{KKI}.  Many
of the theorems proved above will fail in this situation.  For
instance, superdense coding~\cite{Wiesner} provides an example of
a $(2, 2)$ threshold scheme where each share is a single qubit,
but the secret is two classical bits: the four Bell states
$|00\rangle \pm |11\rangle$, $|01\rangle \pm |10\rangle$ encode
the four possible 2-bit numbers, and for all four states, each
qubit is completely random.  This $(2,2)$ scheme is a pure state
scheme, yet does not satisfy corollary~\ref{thm:pure}, and the
share size is smaller than the size of the secret.  Neither is
possible for a purely classical scheme or for a purely quantum
scheme.  Another difference is that there is no rule against
copying classical data, so, for instance, $(k,n)$ threshold
schemes are allowed, even with $k<n/2$.

We can write down conditions for a pure state scheme of this sort
to be secret sharing, along the lines of
theorem~\ref{thm:QSScond}.

\begin{thm}
\label{thm:CSScond} Suppose we have a set of orthonormal states
$|\psi_i\rangle$ encoding a classical secret.  Then a set $T$ is
an unauthorized set iff
\begin{equation}
\langle \psi_i | F | \psi_i \rangle = c(F)
\label{eq:noinfo}
\end{equation}
(independent of $i$) for all operators $F$ on $T$.  $T$ is
authorized iff
\begin{equation}
\langle \psi_i | E | \psi_j \rangle = 0\ \ (i \ne j)
\label{eq:disting}
\end{equation}
for all operators $E$ on the complement of $T$.
\end{thm}

Note that only the basis states $|\psi_i\rangle$ appear in
Theorem~\ref{thm:CSScond}, whereas in Theorem~\ref{thm:QSScond},
the condition had to hold for all $|\psi\rangle$ in a Hilbert
space.  This is the source of the difference between classical and
quantum secrets --- the former hides just a set of orthogonal
states, while the latter hides all superpositions of those states.

\bigskip \noindent
{\bf Proof:} On an unauthorized set, we should be able to acquire
no information about which state $|\psi_i\rangle$ we have.  This
is precisely condition~(\ref{eq:noinfo}).  On an authorized set,
we need to be able to correct for the erasure of the qubits on the
complement.  This is equivalent to being able to distinguish the
state $|\psi_i\rangle$ from the state $|\psi_j\rangle$ with an
arbitrary operator applied to the complement of $T$.  That is, it
is equivalent to condition~(\ref{eq:disting}). \hfill \qed
\bigskip

Note that purely classical secret sharing schemes can be
considered as a particular special case of sharing classical data
with quantum states --- every encoding in a purely classical
scheme is just a mixture of tensor products of basis states.
Purely classical secret sharing schemes are always mixed state
schemes, since classically, there is no way to hide information
without randomness.

Superdense coding provided an example where using quantum data
allowed a factor of 2 improvement in space over any classical
scheme.  It turns out that this is the best we can do.

\begin{thm}
\label{thm:Csize} The dimension of each important share of a
classical secret sharing scheme must be at least as large as the
square root of the dimension of the secret.  The total size of
each authorized set must be at least as large as the secret.
\end{thm}

This means that a $2n$-bit secret requires shares of at least $n$
qubits.

\bigskip \noindent
{\bf Proof:} The proof is quite similar to the proof of
theorem~\ref{thm:Qsize}, which gives the corresponding result for
quantum secret sharing schemes.  We create the quantum state
corresponding to the shared secret 0.  If it is a mixed state
scheme, we include any extra qubits needed to purify it (the
result may not be a secret sharing scheme, however ---
theorem~\ref{thm:mixed} need not hold).  If $S$ is the share under
consideration, and $T$ is an unauthorized set such that $T \cup
\{S\}$ is authorized, give $T$ to Bob, and all the other shares
(including $S$ and the extra purifying qubits) to Alice.

Bob has no information about the secret; $\langle \psi_i | E |
\psi_i \rangle$ is independent of $i$.  Therefore, as in the proof
of theorem~\ref{thm:write}, Alice can perform, without access to
Bob's qubits, a transformation between $|\psi_0\rangle$ (the
current state) and $|\psi_i\rangle$ for any $i$.  Then she sends
the share $S$ to Bob, who now has an authorized set, and can
reconstruct $i$.  We have sent a secret of dimension $s$ using
prior entanglement and the share $S$, which by the bounds on
superdense coding~\cite{CDNT} must therefore have dimension at
least $\sqrt{s}$.  Those bounds also show the size of the channel
plus preexisting entanglement must be $s$, so the size of the full
authorized coalition is at least $s$. \hfill \qed
\bigskip

Note that we used an analogue of theorem~\ref{thm:write} in the
proof.  The general case of theorem~\ref{thm:write} is clearly not
true here: Since the data is classical, we could make two copies
of it.  Then one copy is sufficient to read it, but both are
needed to change it without leaving a trace.  In fact, the version
of the theorem we have used is just the proof that perfect quantum
bit commitment is impossible~\cite{Mayers,LC} --- Bob has no
information about the state, so Alice can change the state to
whatever she likes.

Besides being an interesting result about secret sharing schemes,
this theorem is useful in analyzing other cryptographic concepts.
For instance, it shows that there is no useful unconditionally
secure cryptographic memory protocol, which can only be unlocked
with a key, which we would want to be much smaller than the stored
data.  Such a protocol would be a $(2,2)$ secret sharing scheme,
so the theorem requires that the key be at least half the size of
the data.

Theorem~\ref{thm:Csize} can be easily modified to show that in any
purely classical scheme, each important share must be at least
dimension $s$, not $\sqrt{s}$.  This follows because if Alice and
Bob are just sending classical states back and forth, they need a
channel of dimension $s$ to send the secret rather than dimension
$\sqrt{s}$.  We have already seen one example where this
improvement is achievable using quantum states.

When else can we get this factor of 2 improvement in the number of
qubits per share?  I do not have a full answer to this question.
Certainly for a $(1, n)$ threshold scheme, no improvement is
possible, since each authorized coalition (each single share) must
be as large as the secret.  For many other threshold schemes,
however, an improvement is possible.

\begin{thm}
\label{thm:Cschemes} A $(k, n)$ threshold scheme exists sharing a
classical secret of size $s = p^2$ with one qupit (a
$p$-dimensional quantum state) per share whenever $n \leq 2k-2$,
$p \geq n$, and $p$ is prime.
\end{thm}

Before giving the proof, I will review some basic facts about
quantum and classical error-correcting codes which will be needed
in the construction. A classical linear $[n, k, d]$ code encodes
$k$ bits in $n$ bits and corrects $d-1$ erasure errors. Classical
codes must satisfy the Singleton bound $d \leq n - k + 1$.  A code
$C$ where the bound is met exactly is called an MDS code (for
``maximum distance separable''), and has some interesting
properties.  The dual $C^\perp$ of $C$ (composed of those words
which have vanishing inner product with all words of $C$) is also
an MDS code.  When $C$ is an $[n, k, n-k+1]$ code, $C^\perp$ is an
$[n, n-k, k+1]$ code.  The codewords of the dual code form the
rows of the parity check matrix.  By measuring the parities
specified by the parity check matrix, we can detect errors --- any
parity which is nonzero signals an error. In addition, in an MDS
code, there is a codeword with support exactly on the set $T$ for
any set $T$ of size $d$. See, for instance, chapter~11
of~\cite{MS} for a discussion of MDS codes.

Quantum codes can frequently be described in terms of a
stabilizer~\cite{Gottesman,CRSS}.  The stabilizer of a code is an
Abelian group consisting of those tensor products of Pauli
matrices which fix every quantum codeword.  That is, the codewords
live in an eigenspace of all elements of the stabilizer.  If the
stabilizer contains $2^a$ elements, it is generated by just $a$
elements, and if we have $n$ qubits, the code encodes $n-a$
qubits.  We usually consider the $+1$ eigenspace of the stabilizer
generators, but we could instead associate an arbitrary sign to
each generator.  Tensor products of Pauli matrices have
eigenvalues $\pm 1$, so each set of signs will specify a different
coding subspace of the same size.

Stabilizer codes can be easily generalized to work over higher
dimensional spaces~\cite{Knill}.  We replace the regular Pauli
matrices with their analogs for $p$-dimensional states $X:
|j\rangle \mapsto |j+1\rangle$, $Z: |j\rangle \mapsto \omega^j
|j\rangle$, and powers and products of $X$ and $Z$ (arithmetic is
now modulo $p$, and $\omega = \exp (2 \pi i/p)$).  The eigenvalues
of $X$, $Z$ and their products and tensor products are powers of
$\omega$, so instead of associating a sign with each generator of
the stabilizer, we should instead associate a power of $\omega$.

There is a standard construction, known as the CSS
construction~\cite{CS,Steane}, which takes two binary classical
error-correcting codes and produces a quantum code.  This
construction generalizes easily to qupits.  Take the parity check
matrix of the first code $C_1$ and replace $j$ with $X^j$,
interpreting the rows as generators of the stabilizer.  Take the
parity check matrix of the second code $C_2$ and replace $j$ with
$Z^j$, again interpreting rows as generators of the stabilizer.
The stabilizer must be Abelian --- this produces a constraint on
the two classical codes, namely that $C_2^\perp \subseteq C_1$. If
$C_1$ is an $[n, k_1, d_1]$ code and $C_2$ is an $[n, k_2, d_2]$
code, the corresponding CSS code will be an $[[n, k_1 + k_2 - n,
\min\{d_1, d_2\}]]$ quantum code.

Now consider the classical polynomial code $D_r$ whose coordinates
are $(f(\alpha_1), \ldots, f(\alpha_n))$.  $\alpha_1, \ldots,
\alpha_n$ are $n$ distinct elements of ${\mathbb Z}_p$ (recall
that $p \geq n$), and $f$ runs over polynomials of degree up to
$r$.\footnote{For an appropriate choice of the $\alpha_i$s, $D_r$
is a Reed-Solomon code or an extended Reed-Solomon code.} There
are $r+1$ coefficients to specify $f$, so $D_r$ encodes $r+1$
pits. Given the function evaluated at $r+1$ locations, we can use
polynomial interpolation to reconstruct the polynomial. In other
words, even if $n-(r+1)$ coordinates of the code are missing, we
can reconstruct the $r+1$ coefficients specifying the polynomial.
Thus, this is an $[n, r+1, n-r]$ classical code --- an MDS code.
Also note that $D_r \subset D_{r+1}$.

The codes $D_r$ provide good examples of purely classical secret
sharing schemes~\cite{Shamir}.  If we choose the first $r$
coefficients of the polynomial at random, any set of just $r$
coordinates will contain no information about the remaining
coefficient, so we get an $(r+1, n)$ threshold scheme.  Applying
the CSS construction to the codes $D_r$ and
$D_{r-1}^\perp$~\cite{AB1,AB2} similarly produces good examples of
quantum secret sharing schemes~\cite{QSS}.

With this background, we are now ready to tackle the construction.

\bigskip \noindent
{\bf Proof of Theorem~\ref{thm:Cschemes}:}  We will produce a
class of secret sharing schemes which use one qupit for each share
and encode two classical pits, whereas any purely classical scheme
could only encode one pit.  We will use the classical codes $D_r$
to create $p^2$ related CSS quantum codes with certain useful
properties.  The secret sharing scheme will encode the $p^2$
classical states as the mixture of all states in the corresponding
code from this family.

\bigskip \noindent
{\bf Lemma:} {\sl The parity check matrix for the code $D_{r-1}$
includes a row $R$ such that for any set of $r+1$ coordinates,
there is a linear combination of rows of $D_{r-1}$ with support
exactly on that set of coordinates.  $R$ appears in the linear
combination with coefficient $1$.  Similarly, the dual code
$D_s^\perp$ has, in its parity check matrix, a row $S$ which
appears with coefficient $1$ in a linear combination with support
on any given set of $n-s$ coordinates.}
\bigskip

For instance, we can take $n=4$, $r=2$, $s=1$, $p=5$.  $D_1$ has
generator matrix
\begin{equation}
G = \pmatrix{1 & 1 & 1 & 1 \cr 0 & 1 & 2 & 3}
\end{equation}
(generated by polynomials $1$ and $x$), and $D_1^\perp$ has
generator matrix
\begin{equation}
G' = \pmatrix{2 & 4 & 1 & 3 \cr 3 & 0 & 1 & 1}.
\end{equation}
(The parity check matrix of $D_1$ is the generator matrix of
$D_1^\perp$ and vice-versa.)  By subtracting $j$ times the first
row of $G$ from the second row of $G$, we get a vector with
support on the three-element set excluding coordinate $j$.
Similarly, by adding some multiple of the first row of $G'$ to the
second row of $G'$, we can get a vector with support on any three
coordinates.

\bigskip \noindent
{\bf Proof of Lemma:} The codes $D_r$ and $D_s^\perp$ are linear,
so we only need prove the coefficients of rows $R$ and $S$ are
nonzero --- then some rescaling will always give the result with
coefficient 1.

Since $D_r$ is an MDS code of distance $n-r$, its dual is an MDS
code of distance $r+2$. Thus, the parity check matrix of $D_r$
(which is also the generator matrix of $D_r^\perp$) has a linear
combination of rows with support on any set of $r+2$ coordinates,
but no linear combination of rows has weight $r+1$ or less.  Since
$D_{r-1}$ is included in $D_r$, but encodes one fewer pit, the
parity check matrix of $D_{r-1}$ is just the parity check matrix
of $D_r$ with one row $R$ added. That parity check matrix has a
linear combination of rows with support on any set of $r+1$
coordinates. Since no linear combination of rows of $D_r^\perp$
has weight $r+1$, each of the weight $r+1$ linear combinations
must include a component of row $R$.  A similar argument gives the
result for $D_s^\perp$. \hfill \qed
\bigskip

Now suppose we create the CSS code corresponding to the two
classical codes $D_{r-1}$ and $D_s^\perp$.  We require that $s =
n-r-1$, $2r \geq n$.  Then $s < r$, so $D_s \subseteq D_{r-1}$,
and we have a quantum code.  We are given two classical pits $a$
and $b$ to share among $n$ parties.  Assign a phase $\omega^a$ to
the generator $R$ corresponding to row $R$ of $D_{r-1}$ and a
phase $\omega^b$ to the generator $S$ corresponding to row $S$ of
$D_s^\perp$.  All the other generators have phase $+1$.  Create
the density matrix formed by a uniform mixture over states in the
subspace specified by this stabilizer.  There are $p^2$ of these
mixed states.

\bigskip \noindent
{\bf Claim}: {\sl The set of mixed states
described above define a $(k, n)$ threshold scheme encoding 2
classical pits, with $k = r+1 = n-s$.}
\bigskip

For instance, in the case $n=4$, $r=2$, $s=1$, $p=5$, we get the
stabilizers
\begin{equation}
\begin{array}{ccccc}
 \ & X^2 & X^4 & X & X^3
\\ \omega^a & X^3 & I & X & X
\\ \ & Z & Z & Z & Z
\\ \omega^b & I & Z & Z^2 & Z^3
\end{array}
\end{equation}
with $\omega = \exp(2\pi i/5)$.  The claim is that this gives a
$(3, 4)$ secret sharing scheme.

I now proceed to establish the claim, which will prove
Theorem~\ref{thm:Cschemes}.

For any set $T$ of $k$ coordinates, there will be an element $MR$
of the stabilizer with support on that set of coordinates, where
$M$ contains no factors of $R$ or $S$.  This follows from the
lemma:  There is a linear combination $M + R$ of rows of the
parity check matrix of $D_{r-1}$ with support on $T$.  This linear
combination translates to an element of the stabilizer --- the
rows of the parity check matrix become generators of the
stabilizer, addition of two rows becomes multiplication of the
corresponding generators, and scalar multiplication of a row
becomes taking the corresponding generator to the appropriate
power.

Since $MR$ has support on $T$, we can measure its eigenvalue with
access only to $T$.  $M$ is a product of generators which are not
$R$ or $S$, so the state has eigenvalue $+1$ for $M$, and it has
eigenvalue $\omega^a$ for $MR$.  Thus, the eigenvalue of $MR$
tells us $a$. Similarly, there is an element $NS$ of the
stabilizer with support on $T$, with $N$ having no factors of $R$
or $S$.  We can measure the eigenvalue of $NS$, and it tells us
$b$.  Thus, any set of at least $k$ coordinates is an authorized
set.

A particular value of the secret is encoded as a uniform
distribution over states in the stabilizer code described above.
Thus, the density matrix corresponding to the secret is the
projection on the subspace which is left fixed by the stabilizer.
That is,
\begin{eqnarray}
\rho(ab) & = & \prod_i (I + M_i +M_i^2 + \ldots + M_i^{p-1})
\\ & = & \sum_{M \in S} M
\end{eqnarray}
(normalized appropriately).  The $M_i$ are the generators of the
stabilizer $S$.  Assume the appropriate phase is included in $M$
in this sum (this means that if we wish $M$ to have eigenvalue
$\omega$, we include it as $\omega^{-1} M$, which has eigenvalue
$+1$).

Suppose $T$ is a set of $k-1$ or fewer coordinates. The density
matrix of $T$ is the trace of $\rho(ab)$ over the complement of
$T$.  Now, $X$, $Z$, and all nontrivial products of $X$ and $Z$
have trace 0. Thus, the only terms in the expression for
$\rho(ab)$ which contribute to the trace are those coming from $M$
with weight $\leq k-1$. But the parity check matrices for
$D_{r-1}$ and $D_s^\perp$ contain no rows or linear combination of
rows of weight less than $k$.  Thus, the density matrix of $T$ is
just the identity, regardless of the value of $ab$.  Thus, $T$ is
unauthorized, proving the theorem. \hfill \qed

\section*{Acknowledgements}

I would like to thank Richard Cleve, Hoi-Kwong Lo, Michael
Nielsen, and Adam Smith for helpful discussions.

\end{document}